# A Resource Letter on Physical Eschatology

Milan M. Ćirković

Astronomical Observatory Belgrade

Volgina 7, 11160 Belgrade

Serbia, Yugoslavia

e-mail: mcirkovic@aob.bg.ac.yu

This Resource Letter treats the nascent discipline of physical eschatology, which deals with the future evolution of astrophysical objects, including the universe itself, and is thus both a counterpart and a complement to conventional cosmology. While sporadic interest in these topics has flared up from time to time during the entire history of humanity, a truly physical treatment of these issues has only become possible during the last quarter century. This Resource Letter deals with these recent developments. It offers a starting point for understanding what the physical sciences might say about the future of our universe and its constituents. Journal articles, books, and web sites are provided for the following topics: history and epistemology of physical eschatology, the future of the Solar system, the future of stars and stellar systems, the global future of the universe, information processing and intelligent communities, as well as some side issues, like the possible vacuum phase transition and the so-called Doomsday Argument.

## I. INTRODUCTION: WHAT IS PHYSICAL ESCHATOLOGY?

> Prediction is always difficult, especially of the future.
>
> Danish saying, often quoted by Niels Bohr

The notion of *prediction* is central to the entire scientific endeavor. Even if we do not restrict ourselves to the rather extreme idea that the only purpose of scientific theory is the prediction of experimental outcomes, prediction plays a pivotal role in scientific methodology. The Popperian notion of falsifiability is based on a simple and universally presumed property of scientific theories: their capacity for predicting the outcomes of experiments or observations not yet performed. This should be especially true for the physical sciences.

Physical eschatology (henceforth PE) is the most recent expression of the ancient desire of humanity to learn about the future. The word eschatology (*éschato* = last) was used originally in an exclusively religious light, as "any system of religious doctrines concerning last or final matters, as death, judgment, or an afterlife" and "the branch of theology dealing with such matters" (*Random House Webster*). Examples of such eschatological literature include *The Revelation of St. John* and several of the Qumran texts. The physical sciences slowly encroached upon this field, however, and in the last quarter century a respectable astrophysical discipline arose as a consequence both of the improvement of our empirical knowledge of the universe and of the explosive advances in the theoretical techniques of modelling and prediction. Sir Martin Rees first employed the word "eschatology" in an astrophysical context in the title of his pioneering article of 1969 (Ref. 114), and Fred C. Adams and Gregory Laughlin used the term "physical eschatology" to denote the entire field in 1997 (Ref. 69). PE results now are published regularly in authoritative journals such as the *Monthly Notices of the Royal Astronomical Society*, *Reviews of Modern Physics*, *Astrophysical Journal*, *Nature*, *Science*, or *Physical Review*. Popular accounts appear in many other scientific journals and books. In recent years, PE topics have begun to appear, somewhat shyly, in undergraduate and graduate curricula, mostly in conventional astrophysical and cosmological courses, but sometimes as courses in their own right (one of which served to motivate the most comprehensive PE study to date, Ref. 69).

Since the laws of physics do not distinguish between past and future (with minor and poorly-understood exceptions in the field of particle physics), we do not have a *prima facie* reason for preferring "classical" cosmology to physical eschatology in the theoretical domain. This distinction is still strong in minds of physicists and philosophers alike, however, and one of the purposes of this Resource Letter is to demonstrate the fallacy of this human prejudice.

## A. Prehistory

While physical eschatology is a child of the last quarter century, seriously starting with papers by Sir Martin Rees (Ref. 114), Paul C. W. Davies (Ref. 94), John D. Barrow and Frank J. Tipler (Ref. 96), and Freeman Dyson (Ref. 98), some scientific attempts at solving— or at least highlighting— eschatological problems occured earlier. Before the recent, astrophysical period from roughly 1969 to the present, there was a first serious effort in the 1920s to think and rethink the impact of science and technology on our forecasts and predictions.

These early thinkers were particularly influenced by two distinguished British masters of fiction, Herbert G. Wells and Olaf Stapledon. Wells's extrapolations into the far future of human society (also within the Solar system) in *The Time Machine* (1895), and Stapledon's vision of wakes and tides of future civilizations in *Last and First Men* (1931), captured the minds and imaginations of working scientists. One was the physicist J. D. Bernal, who (together with the polymath J. B. S. Haldane) inspired many modern physical eschatologists, notably Freeman Dyson. (In his Jerusalem lectures, Ref. 8, Dyson mentions the curious fact that the previous owner of his copy of Haldane's *Daedalus* was none other than Albert Einstein.) The history of science tends to be streamlined, ignoring numerous false trails and blind alleys on the road to modern knowledge. To give a flavor of the complications on that road before physical eschatology acquired its present form, I list below some of these early essays.

1. **DAEDALUS or Science and the Future**, J. B. S. Haldane (Kegan Paul, Trench, Trubner & Co., London, 1923). (E) Available, thanks to C. R. Shalizi, at http://www.santafe.edu/~shalizi/Daedalus.html. Though dealing more with what might be termed "biological eschatology" than PE, this book is essential for any historical account of thinking about the future.

2. **ICARUS or the Future of Science**, B. Russell (E. P. Dutton, London, 1924). (E) Available, thanks to C. R. Shalizi, at http://www.santafe.edu/~shalizi/Icarus.html. A rejoinder to Haldane; expresses the well-known pessimism of the great philosopher and mathematician about the justified and humanistic use of future science and technology.

3. "The Last Judgment" in **Possible Worlds and Other Essays**, J. B. S. Haldane (Chatto & Windus, London, 1927). (E)

4. **The World, the Flesh and the Devil**, J. D. Bernal (2$^{nd}$ edition, Indiana University Press, Bloomington, 1969; original 1929). (E) Available, thanks to C. R. Shalizi, at

http://www.santafe.edu/~shalizi/Bernal/. A great inspiration of futurologists, prophets, and physical eschatologists since it appeared. The concluding section starts with a phrase appropriate for almost any PE study: "By now it should be possible to make a picture of the general scheme of development as a unified whole, and though each part may seem plausible in detail, yet in some obscure way the total result seems unbelievable."

5.    "The End of the World: from the Standpoint of Mathematical Physics," A. S. Eddington, Nature **127**, 447-453 (1931). (I)

6.    **The Beginning and the End of the World**, E. T. Whittaker (Oxford University Press, Oxford, 1942). (I)

A fine essay describing the genesis, contents and evolution of Haldane's eschatological writings is

7.    "Last Judgment: The Visionary Biology of J. B. S. Haldane," M. B. Adams, Journal of the History of Biology **33**, 457-491 (2000). (E)

Some thoughts about the influence of early futurology on modern PE can be found in Dyson's Jerusalim lectures, which he published as

8.    **Imagined Worlds**, F. J. Dyson (Harvard University Press, Cambridge, Mass., 1998). (E)

A book manifestly *not* belonging to PE, but exercising a continuous influence on more than one modern physical eschatologist is

9.    **The Phenomenon of Man**, Teilhard de Chardin, translated by B. Wall (Collins, London, 1959). (I,A) This posthumously published study of the great and controversial paleontologist and theologian can be recognized as a strong inspiration not only for Tipler's Omega Point theory (Refs. 52, 168, and 174), but also for the entire twentieth-century future-oriented thinking (e.g. Refs. 47, 48, 54, 55, and Sec. **V.A.**).

**B. The epistemological basis of physical eschatology and the philosophy of time**

We obviously do not think in the same way about past and future. We remember the past, but not the future. In other words, we claim to have secure knowledge (memories) of past events, but only vague hunches, at best, of future events. We seem to *feel* the passage of time, as the special moment we call "now" moves from past to future. Many tomes have been devoted to philosophical, psychological, artistic, and even social aspects of this sensation. These are beyond the scope of this Resource Letter, but they should be considered within a broader framework. We are concerned here with epistemological properties, as well as differences (if any) between *prediction* and *retrodiction* in physical science. This is a formidable topic that has been investigated many times in different contexts (especially in thermodynamics and in classical and quantum field theory), and I present here a point of entry into the literature that is likely to be of interest from the PE point of view. A general feature in the development of the sciences seems to be that philosophical considerations play a role mostly in their formative phases (or in periods of crisis or controversy). In the case of physical eschatology, we do seem to be in a rather early part of its formative phase.

Some additional philosophical background can be found in Refs. 47, 49, 98, 135, 144-153, and in Sec. **V.B**.

10. **The Poverty of Historicism**, K. R. Popper (Routledge and Kegan Paul, London, 1957; originally published in Economica, 1944/45). (I) The relevant part of this famous book deals with the severe limitations of prediction in both "hard" and "soft" sciences. Even if we were perfect Laplacian calculators, we would need to know not just what the true laws of physics were, but that our knowledge of these laws was accurate and complete, which could never be determined on empirical grounds. This argument lies at the core of our necessary assumptions in PE.

11. "Indeterminism in Quantum Physics and in Classical Physics," K. R. Popper, British Journal for the Philosophy of Science **1**, 117-133 (1950). (A) Extends and applies arguments similar to those in Ref. 10.

12. "Symmetry of Physical Laws. Part III. Prediction and Retrodiction," S. Watanabe, Reviews of Modern Physics **17**, 179-186 (1955). (A)

13. "Two types of prediction in Newtonian and quantum mechanics," G. Feinberg, D. Z. Albert, and S. Lavine, Physics Letters A **138**, 454-458 (1989). (A) In many senses a companion paper to Ref. 14.

14. "Knowledge of the Past and Future," G. Feinberg, S. Lavine, and D. Albert, The Journal of Philosophy **89**, 607-642 (1992). (I) A brilliant paper, unfortunately the last in the career of Gerald Feinberg, physicist and philosopher of great insight and originality. Contains an excellent discussion of prediction and retrodiction in both classical and quantum physics, emphasizing and clearing up their limitations: "...in the case of the future motion of planets in the solar system, for example, it would be necessary to include the effects of other stars, and ultimately of other galaxies, if the predictions are to be extended sufficiently far into the future. If some of these distant influences are omitted, then the predictions will become increasingly less accurate as time goes on. Ultimately, since every mass in the universe can influence every other one at least through gravity, a precise description of the future motion of any body would have to include the effects of all other bodies."

15. "Quantum Pasts and the Utility of History," J. B. Hartle, talk presented at *The Nobel Symposium: Modern Studies of Basic Quantum Concepts and Phenomena*, Gimo, Sweden, June 13-17, 1997 (preprint gr-qc/9712001). (A) Relevant context is quantum cosmology; emphasis is placed on retrodiction, but some of the peculiarities of quantum *vs.* classical prediction are also considered.

16. "The Far, Far Future," J. D. Barrow, invited talk at the Symposium "Far-Future Universe: Eschatology from a Cosmic Perspective" (Ref. 57). (I) Apart from a brief history of scientific predictions, this review article contains a wealth of general epistemological issues.

Some of the issue which come into focus most sharply in PE are treated in a praiseworthy antology of Leslie:

17. **Modern Cosmology and Philosophy**, edited by J. Leslie (Prometheus Books, New York, 1998). (I,A)

In a sense, all studies on the foundations and origin of the second law of thermodynamics are relevant to PE, since our perception of the entropy gradient is precisely what enables us to make any prediction at all (indeed, makes our universe *predictable*). The necessity of the entropy gradient for our existence as intelligent creatures was first pointed out by Henri Poincaré (Ref. 18), and later by Norbert Wiener (Ref. 19). The prevailing PE notion in this context for the last half of the 19th and most of the 20th century—a seemingly "obvious" consequence of the second law—has been that of *heat death*, which originated with Hermann Helmholz in 1854 (cf. Ref. 5). This state of maximum entropy has been elucidated in countless articles and monographs, and even textbooks. A large part of the PE discourse represents a struggle with this Boltzmannian concept and its implications. From the vast literature on this subject, here are some of the more inspired writings.

18. **The Foundations of Science**, H. Poincaré, (Science Press, Lancaster, 1946). (A)

19. **Cybernetics**, N. Wiener (John Wiley and Sons, New York, 1961). (A)

20. "Thermodynamics, Statistical Mechanics and the Universe," H. Zanstra, Vistas in Astronomy **10**, 23-43 (1968). (A) An interesting review from the pen of the distinguished stellar astrophysicist.

21. "Heat Death in Ancient and Modern Thermodynamics," G. Kutrovátz, Open Systems and Information Dynamics **8**, 349-359 (2001). (A) An interesting review of "two very different solutions to the problem why the observed 'activity' of nature does not contradict the irreversibility of physical processes."

22. "Entropy and Eschatology: A Comment on Kutrovátz's Paper 'Heat Death in Ancient and Modern Thermodynamics'," M. M. Ćirković, Open Systems and Information Dynamics **9**, 291-299 (2002). (A)

Those interested in topics related to PE are recommended also to consult the following Resource Letters (see also Ref. 79):

23. "Resource Letter RC-1: Cosmology," M. P. Ryan, Jr. and L. C. Shepley, American Journal of Physics **44**, 223-230 (1976). (E,I,A)

24. "Resource Letter CPP-1: Cosmology and particle physics," D. Lindley, E. W. Kolb, and D. N. Schramm, American Journal of Physics **56**, 492-501 (1988). (E,I,A)

25. "Resource Letter ETC-1: Extraterrestrial Civilization," T. B. H. Kuiper and G. D. Brin, American Journal of Physics **57**, 12-18 (1989). (E,I,A)

**26.**    "Resource Letter AP-1: The anthropic principle," Yu. V. Balashov, American Journal of Physics **59**, 1069-1076 (1991). (E,I,A)

## C. Journals

   The papers on PE have appeared in various physical, astronomical, multidisciplinary and philosophical journals. There is yet no journal concentrating specifically on PE. The major scientific journals, grouped by frequency of appearance of PE-related articles, are: *Nature*, *Physical Review D*, *Astrophysical Journal*, *Monthly Notices of the Royal Astronomical Society*, *Reviews of Modern Physics*, *Physics Letters B*, *Icarus*, *American Journal of Physics* and *General Relativity and Gravitation*.

   There are not many important philosophy and history of science journals that have published several studies on PE. The majority of methodological discussions of PE are found in monographs; what little has been published in journals may be found in the pages of *British Journal for the Philosophy of Science*, *Observatory*, *Sophia*, *Philo* (online journal at http://www.philoonline.org) and, in recent years, *Journal of Evolution and Technology* (online journal at http://www.jetpress.org/index.html). In addition, the Doomsday Argument (Sec. **V.B.**) has been discussed in major philosophy journals, such as *Mind*, *Philosophical Quarterly*, *Synthese*, and *Inquiry*.

## D. Reviews and popular accounts

   The allure of the future has captured the attention not only of many working scientists but also the general public, including journalists and editors of distinguished popular-scientific journals. Here is a small selection of papers exposing some of the results of PE in a popular form, or reviewing some of the technical papers or books on the topics I discuss below. This section is admittedly the most incomplete one, since any technical result or monograph in this domain during the last quarter century has generated a significant response in popular journals ranging from *Scientific American* to *Time magazine* and the daily press.

27. "Will the Universe Expand Forever?" J. R. Gott III, J. E. Gunn, D. N. Schramm, and B. M. Tinsley, Scientific American **234** (March), 62-79 (1976). (E) Explains the "basic dilemma" of the large-scale PE: will the universe expand forever or recollapse? Argues strongly for the ever-expanding case, and some arguments are still very relevant.

28. "The Future History of the Universe," J. K. Lawrence, Mercury **VII**, no. 6 (November/December), 132-138 (1978). (E)

29. "The Ultimate Fate of the Universe," J. N. Islam, Sky & Telescope **57** (January), 13-18 (1979). (E)

30. "The Future of the Universe," D. N. Page, and M. R. McKee, Mercury **12** (January-February), 17-23 (1983). (E)

31. "Not the end of the world," J. Silk, Nature **304**, 191-192 (1983). (E) A review of the book by Jamal N. Islam (Ref. 46) from the pen of one of the most distinguished contemporary astrophysicists.

32. "[Review of] The Ultimate Fate of the Universe," A. Lawrence, Observatory **103**, 268-269 (1983). (E) Another review of Islam's book (Ref. 46). Contains an interesting philosophical conclusion: "Theories of the once-only future can be tested only by *waiting*. And then—tested by whom? If a scientific test requires a conscious scientist who understands the result, then the only possible *scientific* theory on the future of life is that it survives!"

33. "The Future of the Universe," D. A. Dicus, J. R. Letaw, D. C. Teplitz, and V. L. Teplitz, Scientific American **248** (March), 74-85 (1983). (E) "A forecast for the expanding universe through the year $10^{100}$." Based partially on the research by the same authors in Ref. 102.

34. "The far future of the Universe," J. N. Islam, Endevoar **8**(1), 32-34 (1984). (E) Based on the same material as Refs. 29, 46, and 97 of the same author. Investigates the far future of an open universe.

35. "L'Avenir de l'univers," N. Prantzos, and M. Cassé, La Recherche **15**, 839-847 (1984) (in French). (E)

36. "After the Sun Dies," T. A. Heppenheimer, Omni (August), 37-40 (1986). (E)

37. "The Future History of the Solar System," J. Maddox, Nature **372**, 611 (1994). (E) News-and-views column by the long-standing editor of *Nature*, devoted to uncertainties about the fate of the Earth faced with the post-Main Sequence Solar evolution; compare Sec. **II**.

38. "This Too Shall Pass," M. Szpir, American Scientist **85** (May-June), 223-225 (1997). (E) A review—somewhat jovial—of the seminal paper by Adams and Laughlin (Ref. 69).

39. "The Future of the Universe," F. C. Adams, and G. Laughlin, Sky & Telescope **96** (August), 32-39 (1998). (E) A popular exposition of the research in PE from the pen of its two distinguished protagonists; compare Refs. 53 and 69.

40. "The Great Cosmic Battle," F. C. Adams and G. Laughlin, Mercury **29** (January/February), 10-15 (2000). (E)

41. "Embracing the End: When the Stars Burn Out," F. C. Adams and G. Laughlin, Astronomy **28** (October), 48-53 (2000). (E)

42. "The Galactic Millenium," G. Laughlin and F. C. Adams, Astronomy **29** (November), 38-45 (2001). (E) Compare with Ref. 73.

The conventional approach to eschatological issues is exemplified by the cursory (though not unsympathetic) treatments in general cosmological reviews, such as the following three.

43. "Our Universe and Others," M. J. Rees, Quarterly Journal of the Royal Astronomical Society **22**, 109-124 (Fourth Milne Lecture) (1981). (E) Part of this breathtaking essay is devoted to reviewing Rees's own (closed-cosmologies) and Dyson's (open/flat-cosmologies) PE results.

44. "The Universe—Present, Past and Future," M. S. Longair, Observatory **105**, 171-188 (1985) (The Halley Lecture for 1985). (I) It devotes precious little space to the questions of the future, exemplifying the prevailing (misguided) notion that the cosmological future is somehow less interesting than the past, but it does contain a wonderful remark: "The future of our Universe is a splendid topic for after-dinner speculation."

45. "The Epoch of Observational Cosmology," T. Rothman, and G. F. R. Ellis, Observatory **107**, 24-29 (1987). (I)

**E. Books**

Topics pertaining to PE have gained a disproportionate amount of attention in popular or semi-popular books, in comparison to the volume of the research literature in the field. This is unusual, since in scientific fields a large number of research papers usually appear in print before the first popular expositions. Consider, for instance, research on the cosmic microwave background (CMB) or even on extrasolar planetology, disciplines that bear some similarity to PE. Although we may speculate why these fields are the reverse of PE (I provide a close-to-exhaustive list of research publications on PE in the following sections), one reason may be a cultural bias towards the future in many strands of Western life during the last quarter century (and particularly after the end of the Cold War). Many popular books, however, do devote much space to "classical" cosmological issues; this is natural in light of the relative scarcity of results in PE proper.

46. **The Ultimate Fate of the Universe**, J. N. Islam (Cambridge University Press, Cambridge, 1983). (E) See Refs. 29, 31, 32, 34, and 97.

47. **The Anthropic Cosmological Principle**, J. D. Barrow and F. J. Tipler, (Oxford University Press, New York, 1986). (A) Chapter X of this famous—but fairly controversial—book is devoted to physical-eschatological issues. For a detailed bibliography of reviews and reactions to this book up to 1991, see Ref. 26.

48. **The Omega Point: The Search for the Missing Mass and the Ultimate Fate of the Universe**, J. Gribbin (Bantam, New York and Heinemann, London, 1987). (E)

49. **Infinite in all Directions**, F. Dyson (Harper & Row, New York, 1988). (E)

50. **End: Cosmic Catastrophes and the Fate of the Universe**, F. Close (Simon & Schuster, New York, 1988). (E)

51. **The Last Three Minutes**, P. C. W. Davies (Basic Books, New York, 1994). (E)

52. **The Physics of Immortality**, F. J. Tipler (Doubleday, New York, 1994). (E) Although the single most controversial reference here, this book is otherwise very hard to classify. It expounds a particular cosmological model—of the topologically-closed and recollapsing-universe type—and interprets it in quasireligious terms, which sometimes seem appropriate, but mostly just funny or absurd. For severe criticisms of Tipler's approach, see Refs. 148, 167, 171, 173 and 175.

53. **The Five Ages of the Universe**, F. C. Adams and G. Laughlin (The Free Press, New York, 1999). (E,I) This is a beautiful popular exposition of the crucial specialized article by the same authors on the topic (Ref. 69), enriched with much of the "classical" cosmological lore, in particular, inflationary models and the primordial nucleosynthesis.

54. **The Future of the Universe: Chance, Chaos, God?**, A. Benz (Continuum, New York, 2000). (E) Contemporary astrophysics reviewed from an openly theist viewpoint; part IV deals with PE. See also Refs. 167-176.

55. **Our Cosmic Future: Humanity's Fate in the Universe**, N. Prantzos, translated by Stephen Lyle (Cambridge University Press, Cambridge, 2000) (E) Rather technologically and optimistically oriented survey of the future; chapter 4 deals with PE proper, with plenty of neat excursions into history of science and philosophy.

**F. Conference proceedings**

To date there has been only two conferences devoted mainly to PE. The first was a Symposium that was held in Budapest and Debrecen, Hungary, July 2-6, 1999. Its proceedings have been published as

56. **The Future of the Universe and the Future of Our Civilization**, edited by V. Burdyuzha and G. Khozin (World Scientific, Singapore, 2000). (A)

The second was a conference on the "Far-Future Universe: Eschatology from a Cosmic Perspective," which was held in Rome, Italy, November 7-9, 2000. Its proceedings have been published as

57. **Far-Future Universe: Eschatology from a Cosmic Perspective**, edited by G. F. R. Ellis (Templeton Press, Radnor, 2002). (A)

Another partially relevant meeting was a Symposium that was held in conjunction with the 160th Annual Meeting of the Astronomical Society of the Pacific, at the University of Maryland, College Park, June 26-28, 1995. Its proceedings have been published as:

58. **Clusters, lensing, and the future of the universe**, edited by V. Trimble and A. Reisenegger (ASP, San Francisco, 1996). (A)

## II. FATE OF THE EARTH, THE SUN, AND THE SOLAR SYSTEM

> So I travelled, stopping ever and again, in great strides of a thousand years or more, drawn on by the mystery of the earth's fate, watching with a strange fascination the sun grow larger and duller in the westward sky, and the life of the old earth ebb away.
>
> H. G. Wells, *The Time Machine* (1895)

The most local and "practical" aspect of physical eschatology pertains to the future of our immediate cosmic neighborhood—the Earth and the Solar system. There are several reasons for investigating this issue. First, the Solar system is regarded traditionally as a good approximation to an isolated astrophysical system; exceptions dealing for instance with the influence of the Galactic tides on cometary orbits as a rule are regarded as highly controversial. Second, the timescales for the future evolution of the Solar system are driven essentially by the evolution of the Sun up and off the Main Sequence, which is regarded as a well-established part of stellar evolution. Third, the timescale for the demise of the Sun and Earth (at least as a viable habitat) is significantly shorter than the vast majority of timescales encountered in the works I survey later. Following the premise that a prediction is more precise and persuasive as its temporal locus approaches the present, this should be the "firmest" aspect of PE. These advantages should be weighed against the extraordinary high precision by astrophysical and cosmological standards that is required to decide meaningfully on the course of future events. Thus, the variation of the Solar radius of the order of 1% or less during the thermal pulsations on the asymptotic giant branch (AGB) will decide whether our Earth will be evaporated or not (see, e.g., Ref. 66). Such precision is not encountered very frequently even in present-day observational astronomy, not to mention other domains of PE! Finally, this issue obviously has the most practical significance (if such a thing can be defined for physical eschatology at all; but see Ref. 177) for our hypothetical descendants in the far future. I include Ref. 67 as a testimony that bold and original scientists are already aware of this aspect of future astrophysical evolution, and are able to offer ideas on solving the existential problems it implies.

I consider here only astrophysical aspects of the future of the Solar system; meteorological and geophysical aspects fall outside of the scope of this Resource Letter.

**59.** "Survival of the Earth and the Future Evolution of the Sun," S. C. Vila, Earth, Moon and Planets **31**, 313-315 (1984). (I) On the basis of rather simplistic assumptions, Vila argues that Earth will be destroyed in the solar red giant's envelope. Calculates the amount of mass accreted by Earth from the solar wind.

**60.** "The fate of the Earth in the red giant envelope of the Sun," J. Goldstein, Astronomy and Astrophysics **178**, 283-285 (1987). (A)

**61.** "Advanced stages in the evolution of the Sun," U. G. Jorgensen, Astronomy and Astrophysics **246**, 118-136 (1991). (A) First detailed study of "the solar evolution all the way from the ZAMS [zero-age Main Sequence] to the end of its life as a red giant." Confirms earlier rough conclusions that the radius of the AGB Sun is "surprisingly close" to 1 AU, and thus the fate of Earth hangs in the balance.

**62.** "Our Sun. III. Present and Future," I.-J. Sackmann, A. I. Boothroyd, and K. E. Kraemer, Astrophysical Journal **418**, 457-468 (1993). (A) The crucial detailed astrophysical study of the future solar evolution.

**63.** "The Expected Morphology of the Solar System Planetary Nebula," N. Soker, Publications of the Astronomical Society of the Pacific **106**, 59-62 (1994). (A) Deals mainly with the possible influence of Jupiter on the planetary nebula created by Sun in its AGB phase.

**64.** "The Effects of Post-Main-Sequence Solar Mass Loss on the Stability of Our Planetary System," M. J. Duncan and J. J. Lissauer, Icarus **134**, 303-310 (1998). (A) Discusses long-term stability of planetary orbits taking into account both Solar mass loss and accretion drag exerted on planets.

**65.** "The Frozen Earth: Binary Scattering Events and the Fate of the Solar System," G. Laughlin and F. C. Adams, Icarus **145**, 614-627 (2000). (A) Discusses the fate of Earth in view of random binary scatterings of passing stars; concludes that chances of serious disruption of Earth's orbit prior to effects of Solar evolution are very small, ~$10^{-5}$. However, in that case of ejection into interstellar space, Earth will cool with timescale of ~$10^6$ years, and settle down in a quasi-equilibrium state; life in hydrothermal vents could "continue in largely unperturbed fashion" even then.

66. "On the Final Destiny of the Earth and the Solar System," K. R. Rybicki and C. Denis, Icarus **151**, 130-137 (2001). (A) Investigates thermal pulses during the Solar red giant and AGB phases. Conclusion: "Mercury will evaporate... and Venus will most probably be destroyed as well. The Earth's fate still remains controversial, but according to the existing evolution sequences for solar models, it is likely that our planet will evaporate during the giant stage of the Sun."

67. "Astronomical engineering: a strategy for modifying planetary orbits," D. G. Korycansky, G. Laughlin, and F. C. Adams, Astrophysics and Space Science **275**, 349-366 (2001). (A) When faced with the Solar ascent up the Main Sequence, our remote descendants—if any—will certainly find strategies for survival more efficient than those presented in this paper. However, as a pioneering contribution in this respect, it certainly deserves attention today!

## III. FATE OF STARS AND STELLAR SYSTEMS

### A. Fate of stars and the final mass distribution function

Before the sun and the light, and the moon, and the stars are darkened and the clouds return after the rain.

*Ecclesiastes*, 12:2

Stellar evolution is the pride and glory of theoretical astrophysics. The detail and accuracy of stellar models over a range of three orders of magnitude in masses, and over almost seven orders of magnitude in evolutionary timescales, have become something of a yardstick for the quality of the modelling endeavor. Thus, it is somewhat surprising that, apart from a pile of potentially applicable results, there is relatively modest interest in PE-related problems in this area. In a nice expression used in Ref. 69, we live in the *stelliferous* era in the history of the universe. This era is characterized by active star formation from interstellar matter throughout the disks of spiral galaxies, and possibly in some other, more exotic environments, like cluster cooling flows and galaxy-merger events. Since the recycling of matter through stellar mass-loss and supernovae obviously is not perfect (since the matter is continually being locked in inert remnants at the rate of a few Solar masses per year in a galaxy like the Milky Way), this process necessarily will come to an end. But how long into the future this era will last is still very, very uncertain. Other unsolved problems abound. The concept of the *final mass function* of stars, introduced in Ref. 69, can be defined as precisely as the concept of an initial mass function, and is potentially of similar interest, but so far it has not been investigated much. A lot of work has to be done in this PE subfield.

**68.** "The coldest neutron star," G. Feinberg, Physical Review D **23**, 3075 (1981). (I) "So the temperature of a neutron star cannot drop below 100 K. The prospects of observing such cold stars do not seem very bright until interstellar travel becomes commonplace, if then."

**69.** "A dying universe: the long-term fate and evolution of astrophysical objects," F. C. Adams and G. Laughlin, Reviews of Modern Physics **69**, 337-372 (1997). (A) Together with Refs. 98, 101, and 114, constitutes a set of landmark papers in physical eschatology. Contains a fascinating treasure of results in various aspects of PE, notably the one linked with the fate of stars and galaxies.

**70.** "The End of the Main Sequence," G. Laughlin, P. Bodenheimer, and F. C. Adams, Astrophysical Journal **482**, 420-432 (1997). (A) First detailed modeling of the complete evolution of the lowest-mass stars over their stupendously long timescales.

**71.** "Gravitational demise of cold degenerate stars," F. C. Adams, G. Laughlin, M. Mbonye, and M. J. Perry, Physical Review D **58**, 083003-1/7 (1998). (A) Notices that "the wavefunction of the star will contain a small admixture of the black hole states" that will emit Hawking's radiation.

**72.** "Future of Galaxies and the Fate of Intelligent Beings," M. M. Ćirković, Serbian Astronomical Journal **159**, 79-86 (1999). (I) Considers the duration of the stelliferous era in several simple models with infall, which are consistent with the usual chemical evolution constraints.

**73.** "The Galactic Millenium," P. Hodge, Publications of the Astronomical Society of the Pacific **112**, 1005-1007 (2000) (E) Supposing that one "galactic year" lasts about 100 million years—period of the revolution of the Solar system around the Galactic

center— Hodge reviews predictions for the state of our environment in 100 billion years (see also Ref. 42).

## B. The fate of the larger gravitating systems

Surprisingly little has been written about the future evolution of large-scale density perturbations, in particular those that manifest themselves today as clusters and superclusters of galaxies. The future of the large-scale structure itself is connected tightly to the realistic cosmological model and the exact form of the density perturbation power spectrum. Both issues are still controversial, although we have made great progress on both during the past decade. In particular, after the results from cosmological supernovae surveys began to be published in 1998, cosmology began to converge on the flat, $\Omega_m \approx 0.3$, $\Omega_\Lambda \approx 0.7$, dark-energy dominated model.

**74.** "Orbits of the nearby galaxies," P. J. E. Peebles, Astrophysical Journal **429**, 43-65 (1994). (A) Peebles analyzes dynamics of the Local Group galaxies and discusses a controversial possibility of future collision between our Galaxy and M31.

**75.** "Future Evolution of Nearby Large Scale Structure in a Universe Dominated by a Cosmological Constant," K. Nagamine and A. Loeb, New Astronomy, in press (2002) (preprint astro-ph/0204249). (A)

## C. The fate of black holes

The bright sun was extinguish'd, and the stars
Did wander darkling in the eternal space,
Rayless, and pathless...

Lord Byron, *Manfred* (1816)

Among astrophysical objects of interest to PE, the ones that occupy the most elevated place are black holes. The reason is obvious: their longevity surpasses by far that of any other known astrophysical object. Before the "black hole revolution" of the 1970s, black holes were believed to be eternal, and that once formed they cannot be undone. However, after the discovery of the black-hole evaporation process by Stephen Hawking in 1974, and the elaboration of the new field of black-hole thermodynamics by Hawking, Jacob Bekenstein, Roger Penrose, Robert Geroch, Robert M. Wald, William G. Unruh and others, the finiteness of their lifetimes became known. But their exact fate (especially in light of the information-loss puzzle) is still not completely clear. In any case, their lifetimes are enormous: a black hole of 1 Solar mass will evaporate (at least until it reaches a mass on the order of a Planck mass) in about $10^{65}$ years, and supermassive black holes of galactic mass likely will live about $10^{98}$ years! Eventually, in the ever-expanding universe, as was shown by Fred C. Adams and Gregory Laughlin (Ref. 69), the incredibly weak Hawking radiation will come to dominate the radiation energy density of the universe.

Even stranger problems are posed by the conjecture of Hawking (Ref. 76) that black holes may pose a fundamental obstacle to any kind of long-term prediction. Let us consider a pure quantum state corresponding to a distribution of matter of mass M » $M_{Pl}$ (Planck mass), which collapses under its own weight. The density matrix of such a state is given by $\rho = |\psi\rangle\langle\psi|$, with vanishing entropy $S = -\,\mathrm{Tr}\,(\rho\,\ln\,\rho)$. If M is high enough, the matter will inevitably form a black hole. Subsequently, the black hole will slowly evaporate by the Hawking process, emitting blackbody radiation (which by definition carries out no information). The semiclassical treatment used by Hawking in his discovery of the black-hole evaporation and in all subsequent discussions will certainly break down when the mass of black hole approaches $M_{Pl}$, but what will happen with the information from the initial state still locked in the black hole? This is the puzzle of black-hole information loss. As is well-known, the possibility Hawking himself proposed is that the black hole simply evaporates completely and the information is irreversibly lost. Although this idea remains the simplest and the least problematic answer to the puzzle, it has provoked a lot of controversy, since it implies that the evolution of the complete system (universe plus black hole) is fundamentally non-unitary, and leads to evolution of pure into mixed quantum states.

Among piles of literature on the long-term behavior of black holes, some of the useful points of entry are the following:

**76.** "Black hole explosions?" S. W. Hawking, Nature **248**, 30-31 (1974). (A) The discovery of finite black-hole lifetimes. "The black hole would therefore have a finite time of the order of $10^{71}$ [$M_{Solar}$/M]$^{-3}$ s. For a black hole of solar mass this is much longer than the age of the Universe..."—a strong understatement, indeed.

77.  "Breakdown of predictability in gravitational collapse," S. W. Hawking, Physical Review D **14**, 2460-2473 (1976). (A) The celebrated paper exposing the possible nonunitarity of the evolution of evaporating black holes.

78.  "Is Black-Hole Evaporation Predictable?" D. N. Page, Physical Review Letters **44**, 301-304 (1980). (A)

79.  "Resource Letter BH-1: Black Holes," S. Detweiler, American Journal of Physics **49**, 394-400 (1981). (E,I,A)

80.  "The unpredictability of quantum gravity," S. W. Hawking, Communications in Mathematical Physics **87**, 395-415 (1982). (A)

81.  "Lectures on Black Holes and Information Loss," T. Banks, Nuclear Physics B (Proc. Suppl.) **41**, 21-65 (1995). (A)

82.  "Black holes and massive remnants," S. B. Giddings, Physical Review D **46**, 1347-1352 (1992). (A) One of the best exposition of the massive remnant hypothesis: Hawking evaporation must end at M ~ several $M_{Pl}$, and a stable remnant (sometimes called "cornucopion") remains. These are bound to be important in what Adams and Laughlin dubbed the black-hole era (Refs. 39, 53, and 69).

83.  "The Hawking information loss paradox: the anatomy of a controversy," G. Belot, J. Earman, and L. Ruetsche, British Journal for the Philosophy of Science **50**, 189-229 (1999). (A) Gives an overview of the nonunitarity puzzle.

84.  "Gravitation, thermodynamics and quantum theory," R. M. Wald, Classical and Quantum Gravity **16**, A177-A190 (1999). (A)

Another interesting issue is the interaction between the dark energy (usually exemplified by the cosmological constant) and black holes, which has been the topic of a lively (and so far unresolved) debate, as indicated in the references below.

85.  A cosmological constant limits the size of black holes," S. A. Hayward, T. Shiromizu, and K. Nakao, Physical Review D **49**, 5080-5085 (1994). (A)

86.  "Possible effects of a cosmological constant on black hole evolution," F. C. Adams, M. Mbonye, and G. Laughlin, Physics Letters B **450**, 339-342 (1999). (A)

87.  "Black holes must die," N. Dalal and K. Griest, Physics Letters B **490**, 1-5 (2000). (A)

**88.** "The Life and Times of Extremal Black Holes," F. C. Adams, General Relativity and Gravitation **32**, 2229-2234 (2000). (A) Though we do not expect to encounter them in nature, they may still play an important role in far future, especially in conjuction with the intelligent influences (cf. Refs. 104 and 149).

**89.** "Proton decay, black holes and large extra dimensions," F. C. Adams, G. L. Kane, M. Mbonye, and M. J. Perry, International Journal of Modern Physics A **16**, 2399-2410 (2001). (A)

For additional discussions of relevance to black holes and their future evolution see also Refs. 127, 128, 131, 149 and 151.

## IV. GLOBAL COSMOLOGICAL FUTURE

> Time ends. That is the lesson of the Big Bang. It is also the lesson of the black hole.
>
> John A. Wheeler, *The Lesson of the Black Hole* (1981)

This is the "true" eschatological topic. Modeling the future of the entire universe depends on our choice of the cosmological model. Obviously, there are cosmological models in which PE is trivial. Historically, the most important of these has been the steady-state model of Herman Bondi and Thomas Gold, and Fred Hoyle. In Bondi and Gold's version, the main premise of steady-state cosmology is the *Perfect Cosmological Principle*, which states that the universe is not only homogeneous and istotropic in space, but also homogeneous in time. In such a universe, PE is reduced to the trivial statement that the universe on large scales will remain the same as it is today throughout the temporal limit t → +∞. The qualification "on large scales" is crucial here, since stars, for instance, live and die and their populations are slowly extinguished in essentially the same manner as in any evolutionary cosmology (as discussed in section **III.A.** above); thus, the "local" part of PE is still valid in the steady-state context. The difference is that on scales of galaxies and larger, things stay the same owing to the creation of low-entropy matter out of nothing (or out of the universal field of negative energy density, as in Hoyle's and William McCrea's subsequent elucidations of the steady-state concept). However, during the "great controversy" (Ref. 90) this view has been rejected by almost all cosmologists in favor of—fortunately from the PE point of view—an *evolutionary* picture of the universe. Thus, we may expect that events different from those already seen will occur in the cosmological future.

90. **Cosmology and Controversy**, H. Kragh (Princeton University Press, Princeton, 1996). (A) By far the best and most comprehensive reference for the formative period of modern cosmology (ca. 1930-1970). Contains an excellent discussion of the motivation behind the steady-state cosmology, some of which (e.g. uniformity of the laws of nature) is relevant to PE.

The basic duality presented by the "standard model" of evolutionary cosmology is whether the universe will expand forever or the gravitational pull of matter fields will be strong enough to halt the expansion and turn it into contraction towards the "Big Crunch." In the older literature, one can find equality between eternal expansion and topological openness and, conversely, between recollapse and topological closeness. However, as elaborated by Lawrence Krauss and Michael Turner, dark energy introduces a degeneracy into the cosmological future, which indicates that even a topologically closed universe ($\Omega > 1$) can expand forever in the presence of, say, a positive cosmological constant. Conversely, a topologically open universe can recollapse into the Big Crunch if the dark energy is attractive (e.g. a negative cosmological constant). However, since all observations suggest a *repulsive* form of dark energy, this option is of rather academic interest.

91. "Geometry and Destiny," L. M. Krauss and M. S. Turner, *General Relativity and Gravitation* **31**, 1453-1459 (1999). (I)

The same degeneracy has been studied from somewhat different perspective by Lawrence Ford:

92. "Unstable fields and the recollapse of an open universe," L. H. Ford, *Physics Letters A* **110**, 21-23 (1985). (A)
93. "Does $\Omega < 1$ imply that the Universe will expand forever?" L. H. Ford, *General Relativity and Gravitation* **19**, 325-329 (1987). (A)

## A. The future of the standard cosmological model: the ever-expanding universe

We almost certainly live in an ever-expanding cosmological domain ("universe"). This has followed from the discovery of large dark-energy density (interpreted as either the cosmological constant or quintessence) in 1998. Of course, observational cosmology long ago suggested similar conclusions on the long-term future of the universe, since all surveys of gravitating matter fell short of the critical density for recollapse. I list some references dealing with the future of ever-expanding universes, either topologically open or dominated by dark energy.

**94.** "The Thermal Future of the Universe," P. C. W. Davies, Monthly Notices of the Royal Astronomical Society **161**, 1-5 (1973). (A)

**95.** "Possible Ultimate Fate of the Universe," J. N. Islam, Quarterly Journal of the Royal Astronomical Society **18**, 3-8 (1977). (I)

**96.** "Eternity is unstable," J. D. Barrow and F. J. Tipler, Nature **276**, 453-459 (1978). (A) The first comprehensive survey of PE in the ever-expanding universe, with particular emphasis on the late metric distortions. "In our view the space-time geometry is becoming more and more irregular at very late times. Both pictures envision an asymptotic approach to a state of maximum entropy; our version of the heat death is different because we have included the gravitational entropy."

**97.** "The long-term future of the universe," J. N. Islam, Vistas in Astronomy **23**, 265-277 (1979). (I) Another of Islam's pioneering contributions to our present-day understanding of large-scale PE.

**98.** "Time without end: Physics and biology in an open universe," F. Dyson, Reviews of Modern Physics **51**, 447-460 (1979). (A) This paper is crucial for our present understanding of PE. It describes evolution of an open or flat (matter-dominated) universe with a host of philosophical, epistemological, and information-theoretic asides. Notable is Dyson's analogy of our position in physics and astronomy with that in mathmathics; for him, PE is a physical analogue of Gödel's theorem on the incompleteness of mathematics.

99. "Matter annihilation in the late universe," D. N. Page and M. R. McKee, Physical Review D **24**, 1458-1469 (1981). (A)

100. "Eternity matters," D. N. Page and M. R. McKee, Nature **291**, 44-45 (1981). (A) A companion paper to Ref. 99. Concludes that for the flat Friedman universe "radiation will never completely dominate the density... matter will always be important." The asymptotic ratio of matter-to-radiation density has been calculated to be about 0.60.

101. "Entropy in an Expanding Universe," S. Frautschi, Science **217**, 593-599 (1982). (A) Refutes the more than century-old idea of the "heat death" of the universe, confirming the early intuition of Pierre Duhem that entropy in the cosmological context only can approach its maximum value asymptotically. Thus, there will always be a thermodynamical arrow of time, although the number and intensity of relevant processes will decrease without limit. However, the conclusion does not apply to the models with event horizons (cf. Ref. 104).

102. "Effects of proton decay on the cosmological future," D. A. Dicus, J. R. Letaw, D. C. Teplitz, and V. L. Teplitz, Astrophysical Journal **252**, 1-9 (1982). (A)

103. "Future and Origin of our Universe: Modern View," A. A. Starobinsky, invited talk at the Symposium "The Future of the Universe and the Future of our Civilization" (Ref. 56). (I) "In any branch of science, sure forecasts exist for finite periods of time only, ranging from days in meteorology to millions of years in the Solar system astronomy. So, how can cosmology be an exception from this general rule? Evidently, it can't."

104. "Life, The Universe, and Nothing: Life and Death in an Ever-Expanding Universe," L. M. Krauss and G. D. Starkman, Astrophysical Journal **531**, 22-30 (2000). (A) Concludes, contrary to Dyson, that "assuming that consciousness has a physical computational basis, and therefore is ultimately governed by quantum mechanics, life cannot be eternal."

105. "Can the Universe escape eternal acceleration?" J. D. Barrow, R. Bean, and J. Magueijo, Monthly Notices of the Royal Astronomical Society **316**, L41-L44 (2000). (A)

106. "Dark Energy and the Observable Universe," E. H. Gudmundsson and G. Björnsson, Astrophysical Journal **565**, 1-16 (2001). (A) Future of Λ- and quintessence-dominated models from an observational point of view; complementary to Ref. 129 by the same authors.

107. "Can we predict the fate of the Universe?" P. P. Avelino, J. P. M. de Carvalho, and C. J. A. P. Martins, Physics Letters B **501**, 257-263 (2001). (A)

108. "The Fate of the Accelerating Universe," J.-A. Gu and W.-Y. P. Hwang, Physical Review D, in press (2002) (preprint astro-ph/0106387). (A)

109. "The Long-Term Future of Extragalactic Astronomy," A. Loeb, Physical Review D **65**, 047301-1/4 (2002). (A) Considers the sky in dark-energy dominated cosmological future; compare with Ref. 106. "In contrast to a matter-dominated universe... the statistics of visible sources in a Λ-dominated universe are getting worse with the advance of cosmic time."

110. "Vacuum Decay Constraints on a Cosmological Scalar Field," J. S. Heyl and A. Loeb, Physical Review Letters **88**, 121302-1/3 (2002). (A) Shows that lack of bubbles of collapsing space-time at present constrains the nature of dark energy and makes untenable the cyclic or ekpyrotic models—our Big Bang preceded by Big Crunch of the previous cycle with minimal value of the scalar potential equaling zero.

111. "Future Island Universes in a Background Universe Accelerated by a Cosmological Constant and by Quintessence," T. Chiueh and X.-G. He, Physical Review D **66**, 123518-1/8 (2002). (A)

112. "Is the Universe Inflating? Dark Energy and the Future of the Universe," D. Huterer, G. D. Starkman, and M. Trodden, Physical Review D **66**, 043511-1/6 (2002). (A)

113. "Accelerating Universe and Event Horizon," X.-G. He, Physical Review D, submitted (2002) (preprint astro-ph/0105005). (A)

## B. The future of the standard cosmological model: the recollapsing universe

Recollapsing-universe models have been associated traditionally with topologically-closed models containing a finite amount of matter (those with $\Omega_m > 1$). The inadequacy of this formulation in the general case has been explained above; nonetheless, I list here references treating such recollapsing world-models. Of course, nowadays it seems highly unlikely that a recollapse will occur. Recent observations of both cosmological supernovae and the CMB anisotropies speak strongly against the possibility of recollapse. This is corroborated by estimates of the age of the universe (coupled with recent data on the Hubble constant) and by the general failure to find anything even remotely close to the amount of gravitating matter necessary for recollapse. The references below show that—in sharp contradistinction to the ever-expanding universe—interest in recollapsing models obviously has declined during the past decade (with an exception of the ekpyrotic model of Steinhardt and Turok, admittedly a "special case").

Recollapsing universes are distinguished by possessing only finite physical time in the future, which may obviate other eschatological results. For instance, if the universe is topologically closed by a large margin (say $\Omega_m \sim 2$), the maximal future time is of the order of $10^{11}$ years, so that processes like Hawking's evaporation of black holes of stellar mass will never occur.

A special case of recollapsing universes which has been quite popular during the 20th century are oscillating models in which the universe passes through a series (allegedly infinite, but see Ref. 130) of expansion and contraction cycles. These models, like the classical steady-state model, blur the difference between past and future, and thus are of only limited interest from the physical eschatological point of view. However, I include here some of the literature dealing with them, both for the sake of completeness and because of the great historical role they played in generating interest in cosmology.

**114.** "The collapse of the universe: an eschatological study," M. J. Rees, Observatory **89**, 193-198. (1969). (I) The pioneering PE study, starting the entire field and coining a new meaning for the old word.

**115.** "Singularities in Cosmology," R. Penrose, in M. S. Longair, ed., **Confrontation of Cosmological Theory with Observational Data** (IAU, D. Reidel Publishing Co., Boston, 1974), 263-272. (I)

**116.** "Speculation on cosmological bounce," M. Bailyn, Physical Review D **15**, 957-964 (1977). (A)

**117.** "General relativity, thermodynamics, and the Poincaré cycle," F. J. Tipler, Nature **280**, 203-205 (1979). (A) Shows the impossibility of "eternal return," i.e., Poincaré recurrence in the cycles of the closed universe governed by general relativity.

**118.** "Gravitational bounce," K. Lake and L. A. Nelson, Physical Review D **22**, 1266-1269 (1980). (A)

**119.** "Phase transitions and dynamics of the universe," V. Petrosian, Nature **298**, 805-808 (1982). (A) "The restoration of symmetry at grand unification in a closed contracting

Robertson-Walker universe could slow down and halt the contraction, causing the universe to bounce and avoid the singular state or the big crunch."

120. "The impossibility of a bouncing universe," A. H. Guth and M. Sher, Nature **302**, 505-506 (1982). (A) A criticism of Petrosian (Ref. 119) with respect to the possibility of a bounce.

121. "Reply by Vahé Petrosian," V. Petrosian, Nature **302**, 806-807 (1982). (A) Reply to Guth and Sher, Ref. 120.

122. "Acceleration and dissolution of stars in the antibang," E. R. Harrison, in G. O. Abell and G. Chincarini, eds., **Early Evolution of the Universe and Its Present Structure** (IAU, D. Reidel Publishing Co., Boston, 1983), 453-455. (A) "Antibang" is Harrison's preferred term for the Big Crunch.

123. "Black holes and the fate of a closed universe," D. Kazanas, in G. O. Abell and G. Chincarini, eds., **Early Evolution of the Universe and Its Present Structure** (IAU, D. Reidel Publishing Co., Boston, 1983), 331. (A)

124. "Thermodynamics and the end of a closed Universe," S. A. Bludman, Nature **308**, 319-322 (1984). (A)

125. "A place for teleology?" W. H. Press, Nature **320**, 315-316 (1986). (I) Criticism of Barrow and Tipler's book (Ref. 47), including its PE aspects.

126. "Achieved spacetime infinity," F. J. Tipler, Nature **325**, 201-202 (1987). (I) Reply to the criticism of Press, dealing explicitly with the history-laden issue of whether it is meaningful to state that an actual infinity of events occur before the final singularity.

127. "Black holes and structure in an oscillating universe," W. C. Saslaw, Nature **350**, 43-45 (1991). (A) "If black holes exist in the contracting phase of a closed universe, they will give rise to a pressure and entropy catastrophe. First, the black holes absorb all radiation; then their apparent horizons merge, and coalesce with the cosmological apparent horizon. ...in these oscillating universes containing black holes, the formation of structure, as well as the existence of life, always gets another chance."

128. "Black-hole mergers and mass inflation in a bouncing universe," A. E. Sikkema and W. Israel, Nature **349**, 45-47 (1991). (A) Argues that, contrary to usual considerations, black holes may be states of very *low* entropy. This would circumvent most of the problems with the bouncing closed universe given since Tolman's time.

129. "Cosmological observations in a closed universe," G. Björnsson and E. H. Gudmundsson, Monthly Notices of the Royal Astronomical Society **274**, 793-807 (1995). (A) "Practical" study of observations in the recollapsing universe. "To an observer in a contracting universe, the night sky would present a colourful zoo of cosmological objects, a vast collection of primaries and ghosts, some blueshifted, others redshifted, where apparent brightness, or size, by itself would not be a reliable indicator of distance, even if all objects were intrinsically the same and not evolving with time."

130. "Oscillating universes," J. D. Barrow and M. P. Dąbrowski, Monthly Notices of the Royal Astronomical Society **275**, 850-862 (1995). (A) "If we live in a closed Friedmann universe that has undergone an infinite number of past oscillations, and if there is a positive cosmological constant, then, no matter how small its value, we might expect most likely to be living in the first phase after the oscillations have ceased, which will eventually become dominated by the cosmological constant."

131. "The Ultimate Future of the Universe, Black Hole Event Horizon Topologies, Holography and The Values of the Cosmological Constant," F. J. Tipler in **Relativistic Astrophysics: 20th Texas Symposium**, AIP Conference Proceedings, volume 586 (AIP, Melville, New York, 2001), 769-772. (A)

132. "A Cyclic Model of the Universe," P. J. Steinhardt and N. Turok, Science **296**, 1436-1439 (2002). (A)

133. "Cosmic Evolution in a Cyclic Universe," P. J. Steinhardt and N. Turok, Physical Review D **66**, 126003-1/20 (2002). (A)

A colorful astrophysical process clearly relevant for late stages of a recollapsing universe is modelled in:

134. "The evolution of irradiated stars," C. A. Tout, P. P. Eggleton, A. C. Fabian, and J. E. Pringle, Monthly Notices of the Royal Astronomical Society **238**, 427-438 (1989). (A)

## C. The future of exotic or nonstandard cosmological models

Models different from standard Friedmann models also have been considered from the point of view of their future evolution. A somewhat peculiar example, which I list here for the sake of completeness, is the famous recollapsing model of Thomas Gold, in which the arrow of time reverses with the reversal of expansion:

**135.** "The Arrow of Time," T. Gold, American Journal of Physics **30**, 403-410 (1962). (I)

**136.** "Will entropy decrease if the Universe recollapses?" D. N. Page, Physical Review D **32**, 2496-2499 (1985). (A) Criticizes the Gold universe, as well as Hawking's support for it; ends with: "Actually, it would not be surprising if the relative probability of our being in the expanding phase is much closer to unity, because this phase is predicted to last an arbitrarily long time, and hence during the subsequent recollapse all stars may have burned out and there may not be much around except for large black holes continually coalescing."

**137.** "Time-symmetric cosmology and the opacity of the future light cone," P. C. W. Davies and J. Twamley, Classical and Quantum Gravity **10**, 931-945 (1993). (A)

**138.** "Observation of the Final Boundary Condition: Extragalactic Background Radiation and the Time Symmetry of the Universe," D. A. Craig, Annals of Physics **251**, 384-425 (1996). (A) The most detailed analysis so far of the time-symmetric cosmological models. "On the dual grounds of theory and experiment, it therefore appears unlikely that we live in a time symmetric universe. (A definitive expurgation must await more thorough investigation of at least some of the aforementioned difficulties.)" Craig finds that "[t]his is therefore a demonstration by example that physics today can be sensitive to the presence of a boundary condition in the arbitrarily distant future."

**139.** "Causality in time-neutral cosmologies," A. Kent, Physical Review D **59**, 043505-1/5 (1998). (A)

This model is arguably closer to the steady-state theory from the PE point of view, since it does not tell us anything particularly interesting or new about the future except, of course, the bizarre and superficially counterintuitive situations encountered in the "counter-clock world"—bizarre, that is, from our perspective but completely normal from the perspective of hypothetical contemporary intelligent beings. Some of the other non-standard models with some PE aspects are:

**140.** "An Isothermal Universe," W. C. Saslaw, S. D. Maharaj, and N. Dadhich, Astrophysical Journal **471**, 571-574 (1996). (A) Derives a class of inhomogeneous cosmologies that "may represent the ultimate state of an Einstein-de Sitter universe that undergoes a phase transition caused by gravitational clustering."

141. "Structure and future of the 'new' universe," Ya. B. Zeldovich and L. P. Grishchuk, Monthly Notices of the Royal Astronomical Society **207**, 23ᴾ-28ᴾ (1984). (A)

142. "Optimistic cosmological model," N. S. Kardashev, Monthly Notices of the Royal Astronomical Society **243**, 252-256 (1990). (A) "It is demonstrated that, for a certain type of hidden mass… a positive curvature cosmological model can realize a regime of periodic oscillations of the Universe without approaching singularity or even a steady-state regime... Finally, note that the model mentioned is most optimistic because it does not lead to the extermination of life as a result of the unlimited expansion of the Universe and of a density decrease or collapse to singularity. This statement also may be accepted as part of the Anthropic Cosmological Principle."

143. "Effects on the Structure of the Universe of an Accelerating Expansion," G. A. Baker, Jr., General Relativity and Gravitation **34**, 767-791 (2002). (A) This paper investigates inhomogeneous mass-distributions in the background universe dominated by cosmological constant. "[I]t appears that for larger scale structures composed of galaxies and inter-galactic space, the observed increase in the rate of expansion may be an important feature in determining the size of self-bound gravitating systems. For smaller structures like galaxies, globular clusters, *etc*. other mechanisms are presumably dominant."

**D. Information processing, intelligent beings, and the cosmological future**

Dyson taught us in his seminal paper (Ref. 98) that, "It is impossible to calculate in detail the long-range future of the universe without including the effects of life and intelligence. It is impossible to calculate the capabilities of life and intelligence without touching, at least peripherally, philosophical questions. If we are to examine how intelligent life may be able to guide the physical development of the universe for its own purposes, we cannot altogether avoid considering what the values and purposes of intelligent life may be. But as soon as we mention the words value and purpose, we run into one of the most firmly entrenched taboos of twentieth-century science." The authors listed below have tried to undermine this taboo. Nonetheless, it should be noted that discussions of life and information processing are still on a different footing than predictions of the future evolution of stars, stellar systems, and the physical universe. The laws of physics are relatively well-known, and even battle-tested. We are still trying to figure out the basic definitions of life, and are far from having a deep, predictive theory of life and intelligence. In spite of this limitation, however, progress can be made and the battle is still raging.

**144.** "Cosmological limits on computation," F. J. Tipler, International Journal of Theoretical Physics **25**, 617-661 (1986). (A) The basic paper on the crucial link among astrophysical evolution, information theory, and intelligent communities.

**145.** "Life after inflation," A. D. Linde, Physics Letters B **211**, 29-31 (1988). (A) In the very first sentence the author states that "one of the main purposes of science is to investigate the future evolution of life in the universe"; concludes that our cosmological domain probably will evolve into an exponential black hole containing inflationary regions inside on huge timescales of ~$10^{10000}$ years! Suggests a "moving" strategy for indefinite survival of intelligent species.

**146.** "World as system self-synthesized by quantum networking," J. A. Wheeler, IBM Journal of Research and Development **32**, 4-15 (1988). (I) This beautifully written paper expounds Wheeler's celebrated notion of the participatory universe; there are several passages of relevance to PE, for instance: "Minuscule though the part is today that such acts of observer-participancy play in the scheme of things, there are billions of years to come. There are billions upon billions of living places yet to be inhabited. The coming explosion of life opens the door to an all-encompassing role for observer-participancy: to build, in time to come, no minor part of what we call *its* past—*our* past, present and future—but this whole vast world."

**147.** "The ultimate fate of life in universes which undergo inflation," F. J. Tipler, Physics Letters B **286**, 36-43 (1992). (A) Criticizes Linde's optimism (cf. Ref. 145) as far as survival of life and intelligence in (chaotic) inflationary universes.

**148.** "Life at the End of the Universe?" G. F. R. Ellis and D. H. Coule, General Relativity and Gravitation **26**, 731-739 (1994). (I) A critical comment on Ref. 147.

**149.** "Possible Implications of the Quantum Theory of Gravity," L. Crane (1994), preprint hep-th/9402104. (E) Expounds what the author calls the meduso-anthropic principle – advanced civilizations creating black holes as a way of proliferating universes in Smolin's manner! "Although it has been generally believed by people with a scientific frame of mind that human life and history take place within the rule of physical law, it has generally been assumed that the relationship between the specific laws of physics and human events was complex and accidental. This has, in fact, placed science in conflict with the otherwise dominant current of Western (and by no means only Western) thought."

**150.** "Can the Universe create itself?" J. R. Gott, III and L.-X. Li, Physical Review D **58**, 023501-1/43 (1998). (A) Opening sections of this remarkable paper briefly consider fates of various cosmological models from the point of view of quantum cosmologies. Section X deals with "baby universes" and possible role of advanced intelligent communities in creating them. Contains one of the best relevant bibliographies.

**151.** "Eternal inflation, black holes, and the future of civilizations," J. Garriga, V. F. Mukhanov, K. D. Olum, and A. Vilenkin, International Journal of Theoretical Physics **39**, 1887-1900 (2000). (A) Considers in detail the problem of information transmission from one inflating region to another; concludes that obstacles (mainly in the form of quantum-energy conditions) are formidable, but that there still is room for the total number of civilized regions in the branching tree of universes to be infinite.

**152.** "The Physics of Information Processing Superobjects: Daily Life Among the Jupiter Brains," A. Sandberg, Journal of Transhumanism **5** (now Journal of Evolution and Technology, at http://www.jetpress.org/volume5/Brains2.pdf), 1-34 (2000). (A) Analyses specific information technologies available to far-future human or advanced extraterrestrial civilizations; many issues are related to PE, which is explicitly considered in §8.4.

**153.** "Cosmological Constant and the Final Anthropic Hypothesis," M. M. Ćirković and N. Bostrom, Astrophysics and Space Science **274**, 675-687 (2000). (I) Reformulates the Final Anthropic Principle of Barrow and Tipler (Ref. 47) into a serious hypothesis about the physical universe. The authors investigate the chances of such a Final Anthropic Hypothesis being true in the realistic cosmological model, dominated by cosmological constant.

**154.** "Ultimate physical limits to computation," S. Lloyd, Nature **406**, 1047-1054 (2000). (A) Although it does not explicitly address PE issues, this paper is important for Lloyd's bold speculations on the future computing technologies, as well as on the computing capacities of black holes. Compare Refs. 144, 152, and 155.

**155.** "On The Maximal Quantity Of Processed Information In The Physical Eschatological Context," M. M. Ćirković and M. Radujkov, Serbian Astronomical Journal **163**, 53-56 (2001). (I)

**156.** "The Ultimate Fate of Life in an Accelerating Universe," K. Freese and W. H. Kinney, Astrophysical Journal, in press (2002) (preprint astro-ph/0205279). (A) Compare to Refs. 98 and 104. Attempts to salvage some of the optimism of the former, arguing that in models going beyond the simplest accelerating expansion, the Dysonian hybernation method might be feasible, in spite of the conclusions of Ref. 104.

### E. Vacuum decay in the future and other quantum-field apocalypses

A small industry has grown up around the notion of a possible future vacuum phase transition. This is not only an eschatological issue in the most literal sense, but it also is connected with the topic of technological development and the capacities of intelligent communities, since the basic idea is that such communities may trigger the phase transition (presumably unwittingly) by conducting very high-energy physical experiments. Although admittedly smacking of science fiction, this idea has been taken seriously even by high-level administrators of modern particle-accelerator laboratories (Ref. 165)! The reason is easy to understand: even if the chance of such an occurence is exceedingly small, its catastrophical ecological impact is incomparably greater than any other conceivable threat, so it deserves close scrutiny. Topics usually investigated together with the vacuum phase-transition threat are the accidental production of strangelets or even mini black-holes in high-energy experiments.

**157.** "Gravitational Effects on and of Vacuum Decay," S. Coleman and F. De Luccia, Physical Review D **21**, 3305-3315 (1980). (A) Classical paper, always quoted in connection with vacuum phase transition at late cosmological times.

**158.** "Is our vacuum metastable?" M. S. Turner and F. Wilczek, Nature **298**, 633-634 (1982). (A)

**159.** "How stable is our vacuum?" P. Hut and M. J. Rees, Nature **302**, 508-509 (1983). (A) First mention of the possibility that the vacuum phase transition may be induced by high-energy physics experiments; rejects the idea for foreseeable human technologies on the basis of comparison with natural cosmic-ray interactions.

160. "Cosmic-ray induced vacuum decay in the Standard model," M. Sher and H. W. Zaglauer, Physics Letters B **206**, 527-532 (1988). (A)

161. "Comment on 'Slightly massive photon,'" M. Sher, Physical Review D **39**, 3513-3514 (1989). (A) Contains a brief discussion of possible phase transition at late cosmological times.

162. "The environmental impact of vacuum decay," M. M. Crone and M. Sher, American Journal of Physics **59**, 25-32 (1991). (I)

163. "Will relativistic heavy-ion colliders destroy our planet?" A. Dar, A. De Rújula, and U. Heinz, Physics Letters B **470**, 142-148 (1999). (A)

164. "Problems with empirical bounds for strangelet production at RHIC," A. Kent (2000), preprint hep-ph/0009130. (A)

165. "Review of speculative 'disaster scenarios' at RHIC," R. L. Jaffe, W. Busza, F. Wilczek, and J. Sandweiss, Reviews of Modern Physics **72**, 1125-1140 (2000). (A) An officially commissioned study of possible hazardous scenarios dealing with inducing vacuum phase transitions or strangelet production at energies available to the new Brookhaven heavy ion collider.

166. "A critical look at catastrophe risk assessments," A. Kent, *Risk*, in press (2002) (preprint hep-ph/0009204). (A) A criticism of the conclusions of Ref. 165 from the "devil's advocate" point of view.

# V. PHILOSOPHY, THEOLOGY, SOCIOLOGY OF THE FUTURE

## A. Theological, philosophical, sociological inferences

As mentioned above, eschatological issues have been understood traditionally as part of the religious, rather than the scientific domain. The transition that occurred mainly in the 1920s (Refs. 1-4) led to the realization that the physical sciences and, ultimately, technology may be used to predict and influence the future on a large scale. This should not be construed, however, as severing all of the links between religious and physical eschatology. The most obvious (although probably not the most instructive) example of the persisting interaction between the two is Tipler's book (Ref. 52), which left a lasting impression on its scientific and philosophical readers, as seen in the references below.

167. "Is Religion Refuted by Physics or Astronomy," Herman Zanstra, Vistas in Astronomy **10**, 1-22 (1968). (I) A companion paper to Ref. 20. Contrasts, among other things, Teilhard de Chardin's eschatological theory to our knowledge about the expanding universe.

168. "The Omega Point as Eschaton: Answers to Pannenberg's Questions for Scientists," F. J. Tipler, Zygon **24**, 217-253 (1989). (I)

169. **Science as Salvation: A Modern Myth and its Meaning**, M. Midgley (Routledge, London, 1992). (I) Gifford Lectures containing an over-skeptical and often-rhetorical critique of Tipler's Omega-point theory.

170. "The Metaethical Alternative to the Idea of Eternal Life in Modern Cosmology," A. V. Nesteruk, Diotima **21**, 70-74 (1993). (E)

171. "The Idea of Eternal Life in Modern Cosmology: Its Ultimate Reality and Metaethical Meaning," A. V. Nesteruk, Ultimate Reality and Meaning **17**, 222-231 (1994). (I)

172. "Piety in the Sky," G. F. R. Ellis, Nature **371**, 115 (1994). (E) A very strong and sometimes unwarranted criticism of Tipler's theory.

173. "The Final Anthropic Cosmology as Seen by Transcedental Philosophy: Its Underlying Theology and Ethical Contradiction," A. V. Nesteruk in Studies in Science and Theology, Volume **5**: The Interplay Between Scientific and Theological Worldviews, Part I, 43-54 (1997).

174. "There are no limits to the open society," F. J. Tipler, Critical Rationalist **3**, no. 02 (available at http://www.eeng.dcu.ie/~tkpw/tcr/volume-03/), 1-20 (1998). (E) Puts the Omega-point theory in a Popperian context.

175. "Colonising the Galaxies," G. Oppy, Sophia **39**, 117-141 (2000). (E) Another harsh criticism of Tipler's Omega Point theory from a philosophical viewpoint. Uses—rather superficially and unfairly—Tipler's theory as a yardstick for all of physical eschatology.

176. "Physical Eschatology," G. Oppy, Philo **4**, (available at http://www.philoonline.org/) no. 2 (2001). (I) Gives arguments to the effect that emotional involvement is inappropriate when dealing with bleak eschatological perspectives of life and intelligence.

**177.** "Cosmological Forecast and Its Practical Significance," M. M. Ćirković, Journal of Evolution and Technology **12** (available at http://www.jetpress.org/volume12/CosmologicalForecast.pdf) 1-14 (2002). (I) Attempts to demonstrate the significance of early decision-making in the context of the entire history of an intelligent community; dependence on the realistic cosmological model is particularly emphasized.

## B. The Doomsday Argument

One of the most intriguing side issues in discussing the future of humanity is the so-called Doomsday Argument, which was conceived (but not published) by the astrophysicist Brandon Carter in the early 1980s, and first expounded in print by John Leslie in 1989 (Ref. 178) and by Richard Gott in 1993 (Ref. 181). The most comprehensive discussion of the issues involved is Leslie's monograph of 1996, *The End of The World* (Ref. 188). The core idea can be expressed through the following urn-ball experiment. Place two large urns in front of you, one of which you know contains ten balls, the other a million, but you do not know which is which. The balls in each urn are numbered 1, 2, 3, 4,... . Now take one ball at random from the left urn; it shows the number 7. This clearly is a strong indication that the left urn contains only ten balls. If the odds originally were fifty-fifty (identically-looking urns), an application of Bayes' theorem gives the posterior probability that the left urn is the one with only ten balls as $P_{post}$ (n=10) = 0.99999. Now consider the case where instead of two urns you have two possible models of humanity, and instead of balls you have human individuals, ranked according to birth order. One model suggests that the human race will soon become extinct (or at least that the number of individuals will be greatly reduced), and as a consequence the total number of humans that ever will have existed is about 100 billion. The other model indicates that humans will colonize other planets, spread through the Galaxy, and continue to exist for many future millennia; we consequently can take the number of humans in this model to be of the order of, say, $10^{18}$. As a matter of fact, you happen to find that your rank is about sixty billion. According to Carter and Leslie, we should reason in the same way as we did with the urn balls. That you should have a rank of sixty billion is much more likely if only 100 billion humans ever will have lived than if the number was $10^{18}$. Therefore, by Bayes' theorem, you should update your beliefs about mankind's prospects and realize that an impending doomsday is much more probable than you thought previously.

The following references show clearly that the Doomsday Argument continues to be a highly controversial topic. In addition, it is one that obviously requires a truly cross-disciplinary approach to explore, considering that the authors are both physicists and philosophers.

See also the references describing the possible hazards owing to the vacuum phase transition and similar calamities (Refs. 157-166), which, one suspects, motivated some of the interest of physicists in the Doomsday Argument.

**178.** "Risking the World's End," J. Leslie, Bulletin of the Canadian Nuclear Society **21** (May 1989), 10-15 (1989). (E) The very first exposition of the Doomsday Argument in print.

**179.** "Is the end of the world nigh?" J. Leslie, Philosophical Quarterly **40**, 65-72 (1990). (I)

**180.** "Doomsday – Or: The Dangers of Statistics," D. Dieks, Philosophical Quarterly **42**, 78-84 (1992). (I) First suggestion of what came to be called the "Self-Indication Assumption" as an answer to the Doomsday Argument conundrum; roughly suggests that your existence favors the existence of many observers in the universe.

**181.** "Implications of the Copernican principle for our future prospects," J. R. Gott, Nature **363**, 315-319 (1993). (I) Gott's—rather fragile—version of the Doomsday Argument.

**182.** "Future prospects discussed," S. N. Goodman, Nature **368**, 106-107 (1994). (I)

**183.** "Future prospects discussed," A. L. Mackay, Nature **368**, 107 (1994). (I)

**184.** "Future prospects discussed," P. Buch, 1994, Nature **368**, 107-108. (I)

**185.** "Future Prospects Discussed: Gott Replies," J. R. Gott, 1994, Nature **368**, 108. (I) Gott's reply to criticisms published in Nature (Refs. 182-184) of his version of the Doomsday Argument.

**186.** "Too Soon for the Doom Gloom?" T. Kopf, P. Krtous, and D. N. Page, preprint gr-qc/9407002 (1994). (A) Proves that the Self-Indication Assumption exactly cancels the Doomsday Argument probability shift.

**187.** "Our future in the universe," J. R. Gott, in Ref. 58, pp. 140-151 (1996). (E) Elaboration of Gott's views of Refs. 181 and 185.

**188.** **The End of the World: The Ethics and Science of Human Extinction**, J. Leslie (Routledge, London, 1996). (I) Monograph largely inspired by the Doomsday

Argument, but containing a lot of interesting empirical material on possible threats to humanity.

189. "Doom Soon?" T. Tännsjö, Inquiry **40**, 243-252 (1997). (I)

190. "A Refutation of the Doomsday Argument," K. K. Korb and J. J. Oliver, Mind **107**, 403-410 (1998). (I) Lists several—rather intuitive—arguments against the conclusion of the Doomsday Argument.

191. "How to predict everything: Has the physicist J. Richard Gott really found a way?" T. Ferris, The New Yorker **75** (July 12 1999), 35-39 (1999). (E) A review of Gott's version of the Doomsday Argument.

192. "The Doomsday Argument is Alive and Kicking," N. Bostrom, Mind **108**, 539-550 (1999). (I) A successful reply to Korb and Oliver.

193. "Comment on Nick Bostrom's 'The Doomsday Argument is Alive and Kicking'," K. K. Korb and J. J. Oliver, Mind **108**, 551-553 (1999). (I)

194. "No one knows the date or the hour: an unorthodox application of Rev. Bayes' Theorem," P. Bartha and C. Hitchcock, Philosophy of Science **66**, S339-S353 (1999). (I)

195. "The Shooting-Room Paradox and Conditionalizing on 'Measurably Challenged' Sets," P. Bartha and C. Hitchcock, Synthese **118**, 403-437 (1999). (A)

196. "Comment l'Urne de Carter et Leslie se Déverse dans celle de Hempel," P. Franceschi, The Canadian Journal of Philosophy **29**, 139-156 (1999) (in French). (A) Develops the analogy between Hempel's raven paradox and the Doomsday Argument.

197. "Predicting Future Duration from Present Age: A Critical Assessment," C. Caves, Contemporary Physics **41**, 143-153 (2000). (I) Attempts to refute Gott's version (Refs. 181, 185, 187) of the Doomsday Argument.

198. "The Doomsday Argument, Adam & Eve, UN$^{++}$ and Quantum Joe," N. Bostrom, Synthese **127**, 359-387 (2001). (A) Summarizes causal problems inherent in the underlying assumption of the Doomsday Argument, christened by Bostrom as the Self-Sampling Assumption.

199. "The doomsday argument and the number of possible observers," K. D. Olum, Philosophical Quarterly **52**, 164-184 (2002). (A) Argues for acceptance of the Self-Indication Assumption in anthropic reasoning.

200. "The Doomsday Argument and the Self-Indication Assumption: Reply to Olum," N. Bostrom and M. M. Ćirković, Philosophical Quarterly, in press (scheduled for January 2003). (A) Argues that the Self-Indication Assumption is a poor guideline in dealing with the Doomsday Argument; criticizes Ref. 199.

201. **Anthropic Bias: Observation Selection Effects**, N. Bostrom, (Routledge, New York, 2002). (A) A wonderfully detailed treatment of many facets of anthropic reasoning, including both the Doomsday Argument and the issue of statistical prediction in cosmology (and PE).

202. "A Critique of Two Versions of the Doomsday Argument – Gott's Line and Leslie's Wedge," E. Sober, Synthese, in press (scheduled for early 2003). (A)

# VI. INSTEAD OF CONCLUSIONS

And so some day,
The mighty ramparts of the mighty universe
Ringed round with hostile force,
Will yield and face decay and come crumbling to ruin.

>      Lucretius, *De Rerum Natura* (ca. 50 BC)

With Earth's first Clay They did the Last Man's knead,
And then of the Last Harvest sow'd the Seed:
Yea, the first Morning of Creation wrote
What the Last Dawn of Reckoning shall read.

>      Omar Khayyám, *The Rubáiyát* (ca. 1100)

Some say the world will end in fire;
Some say in ice.

>      Robert Frost, *Fire and Ice* (1920)

No predictions subject to early test are more entrancing than those on the formation and properties of a black hole, "laboratory model" for some of what is predicted for the universe itself. No field is more pregnant with the future than gravitational collapse. No more revolutionary views of man and the universe has one ever been driven to consider seriously than those that come out of pondering the paradox of collapse, the greatest crisis of physics of all time.

Charles Misner, Kip Thorne, and John A. Wheeler, *Gravitation* (1973)

The world of brute matter offers room for great but limited growth. The world of mind and pattern, though, holds room for endless evolution and change. The possible seems room enough.

K. Eric Drexler, *Engines of Creation* (1987)

One of the main purposes of science is to investigate the future evolution of life in the universe.

Andrei Linde, *Inflation and Quantum Cosmology* (1990)

In my view, the future of the universe is as interesting as its past and so I do not understand why there are not many more papers on this topic.

Abraham Loeb, private communication (2001)

What in the world is physical eschatology?

Anonymous referee, rejecting a previous manuscript of the author (2001)

**Acknowledgements.** My foremost thanks go to Roger H. Stuewer, for his kind help, encouragement, and careful editorial work on a previous version of this manuscript. I am also happy to express special gratitude to Vesna Milošević-Zdjelar, Branislav K. Nikolić, Olga Latinović, Milan Bogosavljević, and Saša Nedeljković for their invaluable help in locating several hard-to-find references, as well as for their useful comments. Vladan Čelebonović and the Dutch Embassy in Belgrade kindly enabled obtaining several electronic subscriptions wherein important references were found. The useful suggestions and encouragement of Freeman J. Dyson, Zoran Živković, Fred C. Adams, Nick Bostrom, Sir Martin J. Rees, Petar Grujić, Robert Bradbury, Yuri Balashov, George Musser, Marina Radujkov, Kacper Rafal Rybicki, Srdjan Samurović, Ken D. Olum, Jelena Milogradov-Turin, Ivana Dragićević, John D. Barrow, Srdjan Keča, Mark Walker, Zoran Knežević, and Abraham Loeb are also hereby acknowledged.